\documentclass{article}    

\usepackage{graphicx}

\title{\textbf{Quantum Boundaries in \mbox{Minkowski Space}}}  
\author{Richard Mould\footnote{Department of Physics and Astronomy, State University of New York, Stony Brook,
\mbox{New York} 11794-3800; http://ms.cc.sunysb.edu/\~{}rmould}}  
\date{}    
   
\begin{document}             

\maketitle              

\begin{abstract}

It is claimed in another paper that the collapse of a quantum mechanical wave function is more than invariant, it is
trans-representational.  It must occur along a fully invariant surface.  The obvious surface available for this
purpose is that of the backward time cone of the collapse event as proposed by Hellwig and Kraus.  This collapse is widely believed to
result in paradoxical causal loops that cannot be removed by special relativistic and/or standard quantum mechanical considerations
alone.  However, the paradox is resolved when we apply the qRule foundation theory that is developed in the other paper.  The causal
and temporal orders of state reduction are then found to be in agreement with one another, and the resulting boundaries in Minkowski
space are shown to have a novel architecture that limits the range of a Hellwig-Kraus reduction in space and time. Although these
boundaries have been worked out using the qRules, they should be the same for any foundation theory that treats the collapse of a wave
in an invariant way, and requires that a collapse destroys the possibility of any further influence on itself -- as do the qRules.

\end{abstract}

\section*{Introduction}
The collapse of a wave function is an undeniable feature of individual quantum mechanical systems.  However, the collapse of a state
along a $t$ = constant surface of an arbitrary coordinate system is unbelievable, inasmuch as nature does not recognize a surface that
is so obviously constructed by humans.  For this reason, a foundation theory must provide for the collapse of a wave along an
invariant surface that is independent of coordinate representations.

The collapse of a wave is not just invariant, it is trans-representational; that is, it is independent of any choice of basis states. 
Furthermore, it will collapse along the surface of the backward time cone that is here called a \emph{conic} surface.  This is the
Hellwig and Kraus state reduction \cite{HK} that has been widely dismissed as being causally problematic \cite{AA}.  The supposed
paradox cannot be resolved by relativity and quantum mechanics alone because  a foundation theory is also required.  This will 
govern the collapse and impose constraints that establish an unambiguous temporal and causal order between reductions.  When  that
is done, it is shown below that a `later' collapse cannot causally penetrate the backward time cone of a `former' collapse.  And since
a collapse occurs against a background of countless widely disbursed prior collapses, the spatial-temporal extent of any collapse
in Minkowski space is limited.  

One consequence of a Hellwig-Kraus reduction is that the newly collapsed state must also be the initial state of the next phase of
evolution.  So initial states as well as collapsed states are conically defined.  Therefore the most natural quantum mechanical state
is a function  mapped onto the conic surface of the backward time cone of an event in space-time.  It is shown that the
dynamic principle projects these functions forward to successor conic states whose vertices lie along a specified world line.

\section*{Non-Local Correlations}

\begin{figure}[b]
\centering
\includegraphics[scale=0.8]{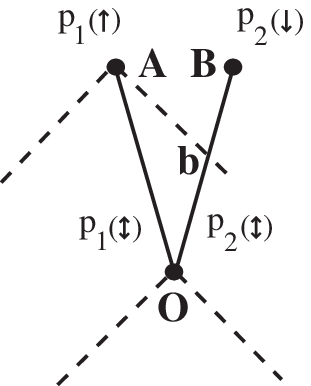}
\center{Figure 1: First non-local reduction}
\end{figure}

Claims regarding a Hellwig-Kraus causal ambiguity are usually advanced by referring to non-local correlations.  Consider a pair of
particles $p_1$ and $p_2$ that are created by a decay  to become correlated in the spin zero state.
\begin{equation}
\psi_0(p_1, p_2) = 2^{-1/2}\{p_1(\uparrow)p_2(\downarrow) - p_1(\downarrow)p_2(\uparrow)\}
\end{equation}
The first particle moves to the left in Fig.\ 1 and the second particle moves to the right.  Initially they both have an uncertain
spin direction as indicated by the double vertical arrow along each path in that figure.  Imagine that the first particle is measured
to have spin-up at an event \textbf{A} in Fig.\ 1, causing a Hellwig-Kraus state reduction.  In that case $p_2$ will go spin-down
when it intercepts the backward light cone of event \textbf{A} at event \textbf{b}.  If $p_2$ is measured  later  at event \textbf{B}
it will of course record spin-down.  The result is the state $p_1(\uparrow)p_2(\downarrow)$.  Capital bold face letters
indicate events that `cause' state reduction like measurements; and lower case bold face letters identify events that are not
vertices  of reduction sites.

Event \textbf{B} also results in a state reduction.  The effect of \textbf{B} on $p_1$ might conceivably  be the
same as the effect of \textbf{A} on $p_2$, where $p_1$ goes spin-up the moment it intercepts the backward
light cone of event \textbf{B} at event \textbf{a} as shown in \mbox{Fig.\ 2}.  So when $p_1$ is later measured at \textbf{A} it
will record spin-up.  The final results are then  the same as those shown in \mbox{Fig.\ 1}, but   this possibility leads to an odd
circularity that is characteristic of a causal loop: 
\textbf{A} causes \textbf{b} causes \textbf{B}, and  \textbf{B} causes \textbf{a} causes \textbf{A}.    

\begin{figure}[h]
\centering
\includegraphics[scale=0.8]{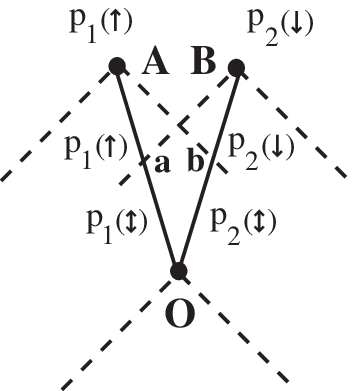}
\center{Figure 2: Second non-local reduction}
\end{figure}

These two particles are shown to follow straight lines even though they are waves spreading out over the indicated paths.  This does
not matter because the rules developed here are not concerned with paths.  We are concerned only with the location of reduction events
like \textbf{A} and \textbf{B} and their associated conic states.  The straight lines in these figures may be thought of as having
heuristic value only.

\section*{Form of qRule Equations } 
The \emph{qRules} are three rules given in another paper that govern the collapse of a wave function \cite{RM1, RM2}.  These rules
generate \emph{qRule equations} that are of the form

\pagebreak

\begin{displaymath}
U(t) = u(t) + u'(t) + ...
\end{displaymath}
where $t$ is \emph{conic time} as opposed to conventional \emph{Lorentz time}.  The Lorentz time at an event \textbf{x} is defined
over a horizontal plane that passes through \textbf{x}, whereas conic time is defined over the conic surface with its vertex at
\textbf{x}. 

It is shown in Appendix I how a conventional Lorentz wave function $\psi(x, y, z, t)$ with its origin at \textbf{x} can be mapped
into a conic wave function $\xi(\mathbf{r}, t)$ with its vertex at \textbf{x}, where \textbf{r} represents a variable that is defined
(in Appendix I) on the conic surface in 3 + 1 space.  We say that the conic wave function $\xi(\mathbf{r}, t)$ mirrors the Lorentz
wave function $\psi(x, y, z, t)$ in a flat  Minkowki space.  The qRule component $u(t)$ is then given by the square modulus of
$\xi(\mathbf{r}, t)$ with its \textbf{r} variable integrated out, making a qRule component a function of $t$ only.  It is also shown
in \mbox{Appendix I} that a conic dynamic principle that mirrors the standard dynamic principle will project the conic function
$\xi(\mathbf{r}, t)$ into successor vertices along the indicated world line.

A qRule equation is always given by a capital $U$ as a function of \mbox{time, but} the components $u(t)$ or $u'(t)$ may be expressed
differently.  For instance, the component $u(t)$ consisting of an atom $a$ and a molecule $m$ can be written \mbox{$u(t) = am(t)$}. 
Each component in a qRule equation is `complete' in that it implicitly contains all the  particles in the universe (Ref.\
3).  The plus sign in a qRule equation always indicates a \emph{discontinuous} quantum jump that is also \emph{irreversible}.

\section*{Decay Application}
The  particle pair $p_1$ and $p_2$ in Eq.\ 1 is assumed to be created at an event \textbf{0} that results from of a decay that begins
at a prior conic time $t_{00}$ along some preferred world line.  If the pair is created by the decay of a composite particle $p_c$, the
corresponding qRule equation is given by 
\begin{equation}
U(t \ge t_{00}) = p_c(t) + \underline{p}_1p_2(t)
\end{equation}
where the $p$'s in this equation are the \emph{qRule values} of the designated particles.  They represent the square modulus of the
conic function of each particle with its independent variables integrated out.  In particular, the component $\underline{p}_1p_2(t)$
is the qRule value of the zero spin wave function $\psi_0$ in Eq.\ 1.  The fact that the first particle goes to the left (in  Fig.\ 1)
and the second particle goes to the right is a distinction that is lost to this component -- because it has been integrated out along
with the spin distinction.  In addition, the component $p_c(t)$ in Eq.\ 2 is the qRule value of the composite particle that produces
the decay particles.  Both of these are  trans-representational components that are derived from wave functions on conic surfaces in
Minkowski space, where $t$ is the conic time referring to these surfaces.  Their evolution is governed by the dynamic principle
operating on the underlying wave function.  Each component is multiplied by an environmental term representing the rest of the
universe, thereby satisfying the requirement that it is `complete'.  The choice of $t$ in these equations does not affect the outcome
as will be demonstrated. 

The second component in Eq.\ 2 is underlined, meaning that it is a \emph{ready} component.  Ready components
are not empirically real, so the component $\underline{p}_1p_2(t)$ in Eq.\ 2 refers to particles that do not yet exist at time $t$ in
that equation.  The non-underlined component $p_c(t)$ is called a \emph{realized} component.   It does exist at time $t$ in that
equation.

The second component $\underline{p}_1p_2(t)$ is zero at time $t_{00}$ and increases in time as probability current flows to it from
the first component $p_c(t)$ in Eq. 2.  The gap between them represented by the + sign is the decay interaction that is
discontinuous and irreversible.  The qRules tell us that the probability current gives  the probability per unit
time that the second component will experience a  stochastic hit.   Only ready components can be stochastically chosen according to
the qRules. If a hit occurs at a time $t_0$ (the time of event \textbf{0}), then the rules tell us that there will be a collapse of
\mbox{Eq.\ 2} to
\begin{equation}
U(t \ge t_0 > t_{00}) = p_1p_2(t)
\end{equation}
where the first component in Eq.\ 2 goes to zero. The wave function associated with Eq.\ 3 is also mapped onto backward time-cone
surfaces -- in this case on vertices  of the successors of event \textbf{0}, where the time variable is the conic time referring to
these vertices.

Equations 2 and 3 reflect the fact that the collapse of a wave function in quantum mechanics is a trans-representational affair. 
 The dynamic principle directs the evolution of a qRule equation, but the qRules govern the
stochastic process that interrupts its continuous flow.   The foundation theory of Ghirardi et. al. incorporates a collapse directly
into the dynamic principle \cite{GPR}, but that theory is not trans-representational.

\section*{Hellwig-Kraus Application}
When the spin-measuring devices $M_1$ and $M_2$ of particle $p_1$ and $p_2$ are later introduced, Eq.\ 3 becomes the qRule equation
\begin{eqnarray}
U(t \ge t_0) =p_1p_2\otimes M_1M_2(t) &+& [\underline{p}_1(\uparrow)M_1]p_2(\downarrow)\otimes M_2(t)  \\ 
&+& \underline{p}_1(\uparrow)[p_2(\downarrow)M_2]\otimes M_1(t)  \nonumber \\
&+& [\underline{p}_1(\downarrow)M_1]p_2(\uparrow)\otimes M_2(t) \nonumber \\
&+& \underline{p}_1(\downarrow)[p_2(\uparrow)M_2]\otimes M_1(t) \nonumber
\end{eqnarray}
where both measuring devices are on standby in the first component of Eq.\ 4, and the four `ready' components (on the right) are zero
at $t_0$.  In the ready component of the first row the first particle engages the spin-measuring device $M_1$ (in square brackets),
and in the second row the second particle engages $M_2$.  The third and fourth rows are similar except that they provide for the
reverse spin measurements. 

Probability current begins to flow from the first component to the ready components in the first and third rows when the first
particle interacts with the measuring device $M_1$ sometime after $t_0$.  Current will begin to flow  to the
ready components in the second and fourth rows when the second particle interacts with $M_2$.  So all four ready components are
exposed to the possibility of a stochastic hit. Figure 3a is a graphic description of \mbox{Eq.\ 4}.  The gray area in that figure
represents the Minkowski region of interaction where the ready components in Eq.\ 4 become non-zero (i.e., where the particle
interacts with the detector).

\begin{figure}[b]
\centering
\includegraphics[scale=0.8]{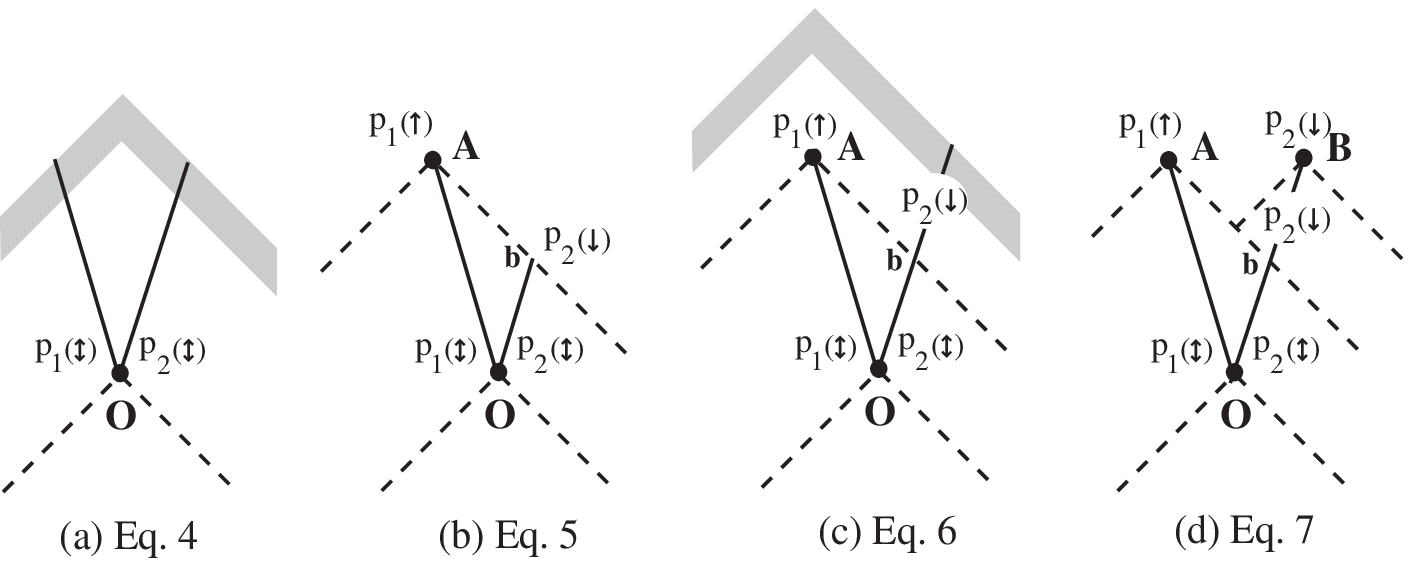}
\center{Figure 3: Non-local reductions}
\end{figure}

Neither one of the world lines in Fig.\ 3a defines the time $t$ in  Eq.\ 4.  That time specifies the chosen orientation (i.e., the
chosen Lorentz observer) in the Minkowski space. Since the probability of any one of these ready states being stochastically 
chosen is independent of the choice of \emph{t},  Eq.\ 4 is invariant under \emph{t}. If another $t$ is chosen, then the world lines
in Fig.\ 3 will be rotated to the left or to the right (as with a Lorentz transformation), but the orientation of the conic surfaces
will be unchanged.  The graphic relationships in Fig.\ 3 will be unchanged, except event simultaneity.

Suppose the first row in Eq.\ 4 is stochastically chosen at a time $t_{A}$.    The
system is then reduced to
\begin{equation}
U(t = t_{A} \ge t_0) = [p_1(\uparrow)M_1]p_2(\downarrow)\otimes M_2(t)
\end{equation}
where $t_{A}$ is the time of the stochastic hit at event \textbf{A} that appears along the world line of $p_1$ (see Fig.\ 3b
representing Eq.\ 5).  It is again understood that the wave functions representing $p_1, p_2, M_1$, and $M_2$ in this figure are
spread out over the conic surface of the vertex of event \textbf{A} in Fig. 3b. Equation 5 not only represents event \textbf{A}, it
also represents the cut-off event \textbf{b}  that is simultaneous with \textbf{A} in conic time.    We make no assumption
as to the world lines of the measuring devices.   

Further evolution carries the resulting conic state  forward  following event \textbf{A} until $p_2$ interacts with
$M_2$.  A new ready state then emerges giving
\begin{eqnarray}
U(t \ge t_{A} > t_0) &=& [p_1({\uparrow})M_1]p_2({\downarrow})\otimes M_2(t)    \\ 
&+& [\underline{p}_1({\uparrow})M_1][p_2({\downarrow})M_2](t) \nonumber
\end{eqnarray}
which is graphically represented in Fig.\ 3c in which the shaded area is the region of the new interaction that gives rise to the
ready component in the second row of Eq.\ 6 when $p_2$ encounters $M_2$.

The evolution of the second particle after event \textbf{b} in Fig.\ 3c repeats part of the evolution that has already occurred in
Fig.\ 3a. The latter part of this evolution involves the interaction $[p_2(\downarrow)M_2]$ between the second particle and its
measuring device.  This part is not empirically significant because its repetition consists in its appearance in both `ready'
components of both Eqs.\ 4 \mbox{and 6}.  An interaction repetition like this would occur even if the time in Eqs.\ 4 \mbox{and 6}
referred to Lorentz time, so it is not a characteristic of a Hellwig-Kraus reduction.  It is a feature of \emph{any} correlated state
reduction.  However, the free particle part of the repetition certainly does  involve the realized component in Eqs.\ 4
\mbox{and 6}, so it does (in principle) have empirical significance. It is also characteristic of a Hellwig-Kraus collapse.  But this
free particle part is not really measurable because it exists independent of a measurement interaction.  

Let a second state reduction occur at event \textbf{B} at time $t_{B}$, giving the final result
\begin{equation}
U(t = t_{B}) = [p_1(\uparrow)M_1][p_2(\downarrow)M_2](t)
\end{equation}
Equation 7 appears in Fig.\ 3d.  

Since the qRules are trans-representational they give us an objective account of what happens.  Equation 4 causally precedes Eq.\ 5,
and that causally precedes Eq.\ 6, and that causally precedes Eq.\ 7.  Causal priority is established by the \emph{formal}
priority among these equations.  A collapse equation comes into existence only when its predecessor disappears.  We state the 
corollary.

\noindent
\textbf{Corollary:} \emph{The evolution of a qRule equation cannot affect the evolution of a previously `collapsed' qRule equation --
since the latter no longer exists.} 

This means that the evolution described in Eq.\ 6 can have no influence on the evolution described in Eq.\ 4; for when Eq.\ 6 is
actively  evolving, Eq.\ 4 \emph{no longer exists}.  Therefore, event \textbf{B} cannot influence event \textbf{A}.  This is
indicated in Fig.\ 3d where the backward time cone of event \textbf{B} is not allowed to `penetrate' the backward time cone of event
\textbf{A}.

\begin{figure}[b]
\centering
\includegraphics[scale=0.8]{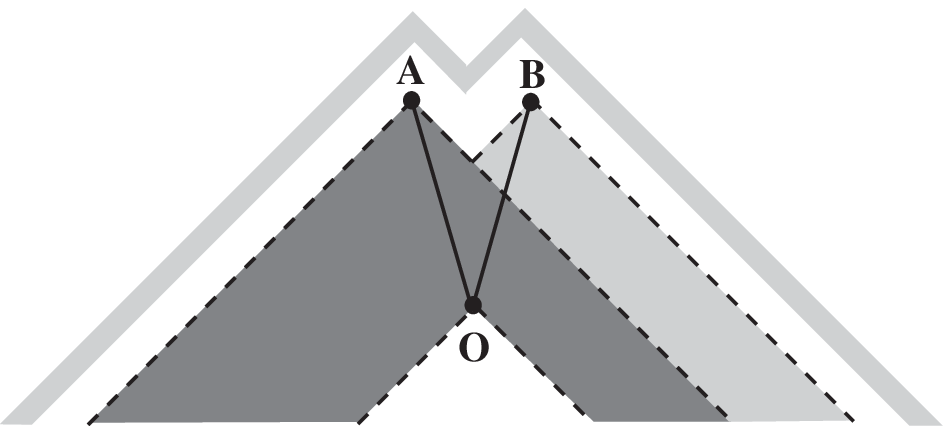}
\center{Figure 4: Causal and temporal order -- non-penetration}
\end{figure}

It is the qRules that govern the causal order of reduction, not  relativity or standard quantum mechanics, and
these rules are perfectly clear concerning priority. This is shown graphically in the Minkowski space of Fig.\ 4.  The more darkly 
shaded area (below \textbf{A} in Fig.\ 4) indicates the region of Minkowski space that evolves according to Eq.\ 4; and the more
lightly shaded area (below \textbf{B}) indicates the region that evolves according to Eq.\ 6, where the lighter area can have no
influence on the darker area because the latter is causally prior.  The reductions in Fig.\ 4 go to infinity on the left and on the
right as  expected of a Hellwig-Kraus collapse.  However, the event \textbf{B} collapse is limited  in that it cannot
penetrate the event \textbf{A} reduction.  This will generally be the case.  The universe is full of prior reductions that
cannot be penetrated by either events \textbf{A} or \textbf{B}; and as a result, both reductions will be limited in the space-time
extent of their influence.  This is not just true of correlated reductions, but will also be true of independent reductions
(see Appendix II).

\section*{Independent Systems}
Following event \textbf{B}, the measuring device $M_1$ (that now includes particle $p_1$) evolves \emph{independent} of the measuring
device
$M_2$ (that now includes $p_2$).  These two systems can be described along two different conic world lines designated $t_1$ and
$t_2$, where $t_1$ begins with the vertex at event \textbf{A}, and a conic time $t_2$ begins with the vertex at event \textbf{B}. 
It is not necessary to introduce these dual times, but doing so helps to emphasize the spatial separation as well as the independence
of the two systems.  The underlying conic function is then of the form $\xi(t_1; t_2)$ which comes about by using two different origins
(on the horizontal $x$-axis) to describe two widely separated systems.  As a result, a single cone spreading over two systems of this
kind is replaced by an envelope that drapes over both -- like the lightly shaded area over events \textbf{A} and \textbf{B} in Fig.\
4.  The qRule equation following event \textbf{B} is then written 
\begin {displaymath}
U(t_1 \ge t_A; t_2 \ge t_B) = [p_1(\uparrow)M_1(t_1) + \underline{i}_1(t_1)][p_2(\downarrow)M_1(t_2) + \underline{i}_2(t_2)]
\end{displaymath}
where $i_1$ is an interaction that is located in some part of the lightly shaded envelope above event \textbf{A} in Fig.\ 4, and
$i_2$ is another interaction that is located in the lightly shaded envelope above event \textbf{B}.  We simplify the equation by
dropping the time dependence in each component
\begin {equation}
U(t_1 \ge t_A; t_2 \ge t_B) = [p_1(\uparrow)M_1 + \underline{i}_1][p_2(\downarrow)M_1 + \underline{i}_2]
\end{equation}
where subscript-1 states are understood to be on the $t_1$ conic surface, and subscript-2 states are understood to be on the $t_2$
conic surface.

Equation 8 might not seem entirely correct according to the qRules. The recipient state $i$ in each bracket of the equation does not
appear to be a complete component; and if that were true, it would not be a ready component.  However, the equation can be written in
the form
\begin{displaymath}
 U(t_1 \ge t_A; t_2 \ge t_B) = p_1(\uparrow)M_1[p_2(\downarrow)M_2 + \underline{i}_2] + \underline{i}_1[p_2(\downarrow)M_2 +
\underline{i}_2]
\end{displaymath}
The brackets in this equation have a square modulus that is constant in time, so the current from the first component comes solely
from $p_1(\uparrow)M_1$ and goes exclusively to $\underline{i}_1$.  Since both components  are clearly complete, the first must be
a realized component and the second is a ready component.  Both \emph{i}'s in Eq.\ 8 can be understood in this way. We therefore
keep the form of Eq.\ 8.  It can be generalized to any number of independent systems in a way that satisfies the qRules.

\vspace{.6cm}

 In 2 + 1 space the mountaintops in Fig.\ 4 do not appear in front or in back of one another.  Rather, they are superimposed on one
another like mountains on a flat terrain.  A new peak will not penetrate an old peak on that terrain because the above corollary
holds for all completed state reductions -- including independent as well as correlated reductions as will be shown in Appendix II. 
These mountaintops all build on top of one another without penetration.

\section*{Another Lorentz Observer}
It is important to understand how Eq.\ 4 and Fig.\ 3 play out relative to a Lorentz frame in which event \textbf{A} occurs
(causally) before event \textbf{B} but appears to follow \mbox{event \textbf{B}} in Lorentz time.  Figure 5a is identical with Fig.\
3a where the regions of measurement interaction for $p_1$ and $p_2$ are indicated by the darkened world lines.  Figure 5b looks at the
same interactions relative to a Lorentz frame in which the second particle is at rest.  This is called the `primed' frame.  

\begin{figure}[b]
\centering
\includegraphics[scale=0.8]{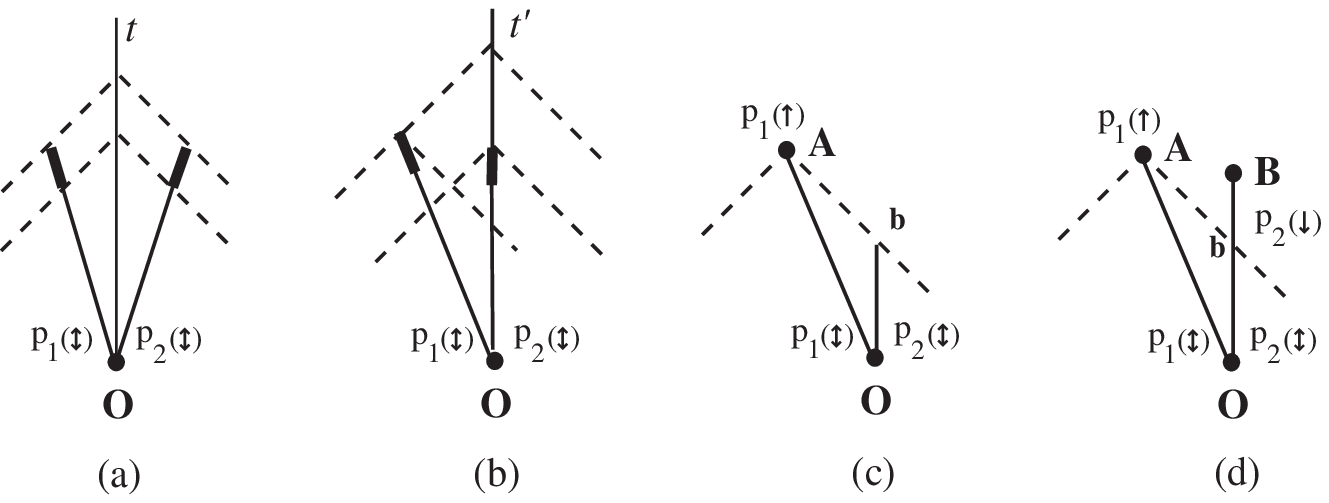}
\center{Figure 5: Reduction in t$^\prime$ frame}
\end{figure} 

Equation 4   has the same form relative to the primed frame  as it does relative to the unprimed frame.  The difference
is only in the way that current flows from the realized component to the four ready components.  In the unprimed frame (Fig.\ 5a)
current flows simultaneously into all four components for as long as it takes for a stochastic hit on the first component (in our
example).  In the primed frame (Fig.\ 5b) current begins to flow into the second particle  \emph{before} it begins to flow into the
first particle -- in  Lorentz  time.  However, this current will run out without the second particle being stochastically chosen,
while current  continues to flow into the first particle.  As a result, the first particle is the first to be stochastically chosen at
event \textbf{A} (in Fig.\ 5c), thereby cutting off the second particle at \mbox{event \textbf{b}}, which is again \emph{simultaneous}
with event \textbf{A} in conic time.  The second particle is subsequently revived at a time that allows it to again interact with the
measuring device $M_2$, thereby allowing it to be stochastically chosen at event \textbf{B} (Fig.\ 5d) \emph{after} event \textbf{A}
in conic time -- although it is \emph{before} \mbox{event \textbf{A}} in Lorentz time. The second particle does not exactly repeat its
prior history.  The first time that it interacts with the measuring device it is not stochastically chosen, but the second time
(following event \textbf{b}) it is stochastically chosen.  The first time the probability of a hit on event \textbf{B} is 0.25, the
second time it is 1.0.  This repetition has no empirical significance because both interactions appear inside ready components.  In
general, stochastic interactions are not empirically real in qRule theory.  Stochastic hits are empirically real.      

Figures 3 and 5 represent two different Lorentz observers, where  the causal  and the  temporal order of events are the same in
both frames.  Only in Lorentz time does there appear to be some disagreement between the causal and temporal orders.  We attribute
this to fact that Lorentz time is given along artificiality defined horizontal planar surfaces.

\section*{Other Foundation Theories}
The structure in Figs.\ 3 and 5 should be the same for any foundational theory that is viewed in an invariant way.  If the
theory in question provides for the collapse of a wave like Eq.\ 1, and if the collapse travels backward over a conic surface in a way
that destroys the possibility of any further influence on itself, then it should produce the same Minkowki architecture that appears
in \mbox{Fig.\ 4}.  For any such theory, in any Lorentz frame, the measurement interaction of the second particle will run its course
without result (in our example) allowing \mbox{event \textbf{A}} to be chosen first. The resulting collapse will re-start the second
particle at an event like
\textbf{b} in Fig.\ 3b that is conically \emph{simultaneous} with \textbf{A}, so event \textbf{B} will occur \emph{after} event
\textbf{A} in conic time.  As in our qRule analysis, it is the reduction theory that governs the casual order, not relativity or
standard quantum mechanics.  The result will be the succession of mountain peaks like those shown in Fig.\ 4, where a peak in
background occurs causally and temporally \emph{after} a peak in foreground, and where both are  limited in space and time by other
reductions.

\section*{Conclusion}
A paradoxical causal loop seems to appear when the collapse of a quantum mechanical wave function is viewed as a Hellwig-Kraus
reduction in Minkowski space.  This paradox cannot be removed by special relativity and/or standard quantum mechanics alone, but
requires the constraints imposed by a suitable foundation theory governing the collapse.  In this paper the qRule foundation theory is
used to resolve this causal difficulty. 

We define conic time $t$ over the surface of a backward time cone of some event \textbf{x}.  A quantum mechanical wave function
$\xi(t)$ is also mapped onto that surface.  This function is used instead of the usual quantum mechanical wave function $\psi$
that is defined on a horizontal surface going through \textbf{x}.  

A Hellwig-Kraus reduction describes two correlated space-like reduction ev-ents \textbf{A} and \textbf{B}, and it is found that the
qRules establish a definite causal priority between them.  There is no \emph{causal loop} or temporal ambiguity between these two
events when the qRules govern the collapse.  The causal and conic temporal orders of events \textbf{A}
and \textbf{B} are the \emph{same}.  The resulting landscape in Minkowski space is shown in Fig.\ 4, where each
mountaintop in that figure is a separate reduction.  The  peak in the background of Fig.\ 4 will occur causally and temporally
\emph{after} the peak in the foreground.  The background reduction cannot `penetrate' or causally influence the interior of the
foreground reduction, as shown in the figure.  Any reduction will take place against a background of countless prior reductions that
cannot be penetrated, thereby limiting its range in space and time.  This will be true even if the reduction is independent (i.e., not
a correlation) as is shown in Appendix II.    

It is my belief that the structure of Fig.\ 4 will be similar for any foundation theory that is
viewed in an invariant way, so long as the theory requires that a collapse destroys the possibility of any further influence on itself
-- as do the qRules.

The dynamic principle is understood to influence the temporal development of a qRule component through the underlying function
$\xi(r, t)$.  This function is defined in  this paper and evolves in time as described in Appendix I.  However,
the qRules govern state reduction and therefore interrupt the dynamic process under the right conditions.  More will be said in a
subsequent paper about the relationship between the dynamic principle, the qRule equations, and the architecture of state reduction in
Minkowski space.

\section*{Appendix I}

\begin{figure}[h]
\centering
\includegraphics[scale=0.8]{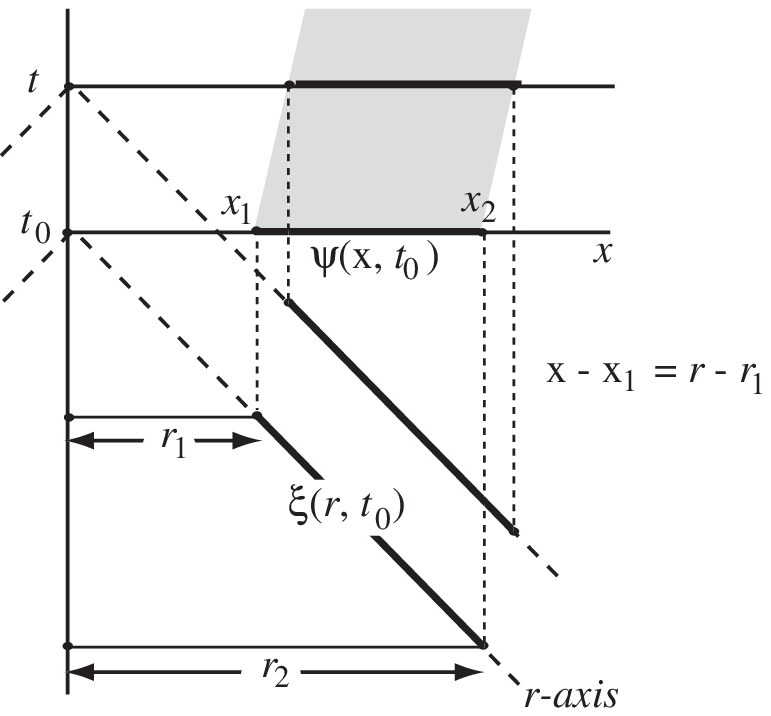}
\center{Figure 6: Mapping onto a conic surface}
\end{figure}

A  particle is initially located at $t_0$ between $x_1$ and $x_2$ in Fig.\ 6.  Its wave function $\psi(x, t)$ advances along the
$+x$-axis occupying the shaded area in flat Minkowski space. This function is mapped onto the dark line on the surface
of the conic section with its vertex at $t_0$.  Each event on that surface has  assigned coordinates $r$ (perpendicular to the
$t$-axis) and $t_0$, so we project $\psi(x, t_0)$ onto the conic surface giving
\begin{displaymath}
\xi(r_2 \ge r \ge r_1, t_0) = \xi(r, t_0)
\end{displaymath}
where $x - x_1 = r - r_1$ holds for all $x$ and $r$.  If $x_1$ is negative, this function can be extended  over the vertex to the
rising side of the conic surface without difficultly. More generally we say that any wave function in flat space that is mapped onto a
conic section with its vertex at
$t$ is given by $\xi(r, t)$.

We want to know how to write the dynamic principle for $\xi(r, t)$.  Assume that it is analogous to the Schr\"{o}dinger equation in
the non-relativistic case, with a dynamic operator $\hat{\mathbf{D}}$ in place of the Hamiltonian.
\begin{equation}
\mathbf{\hat{D}}\xi(r, t) = -i\hbar\partial_t\xi(r, t)
\end{equation}
This equation is  correct if we let

\begin{equation}
\mathbf{\hat{D}} =\frac{\hbar^2}{2m}\partial^2_r
\end{equation}
The function that is mapped onto successive conic surfaces is therefore propagated in time by a dynamic principle that mirrors the
Schr\"{o}dinger equation for a non-relativistic system. Equations 9 and 10 applied to $\xi(r, t)$ will advance it from one conic
surface to another.

The same construction can be applied to each of the four components of a Dirac wave function, in which case the dynamic operator  
$\hat{\mathbf{D}}$ mirrors the Dirac Hamiltonian.  The amplitude in the above equations  then has the required
four components.

In 2 + 1 space the surface is a cone with a vertex time $t$.  A function $\psi$ defined on a horizontal plane with its origin at the
vertex is projected onto the cone's surface. Its $x, y$ coordinates are mapped onto $r_x, r_y$ coordinates that intercept the conic
surface and are perpendicular to the $t$-axis.  The dynamic principle causes these surfaces to evolve along the specified world line.

In 3 + 1 space the incoming `spherical' conic surface will converge on a vertex time $t$.  The particle volume that is simultaneous
with $t$ contains the initial condition $\psi_0(x,y, z, t)$, and each of these values can be mapped onto the incoming spherical
surface when (prior to $t$) the surface passes through each particle part at its time $t$ locaton.  

Our claim is that the $x$-axis is superfluous, and that the only coordinates we need are those associated with a conic surface.  If
given the wave function $\xi(r, t)$ of a particle along the surface with vertex $t_0$ in Fig.\ 7, the dynamic principle will
carry it into all succeeding surfaces like that of $t_1$ in that figure.

\begin{figure}[h]
\centering
\includegraphics[scale=0.8]{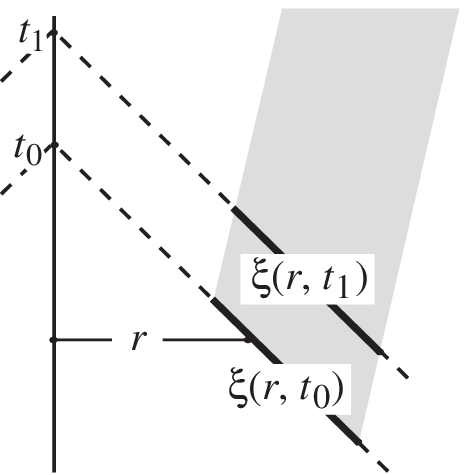}
\center{Figure 7: General conic wave}
\end{figure}

The `real' location of the particle -- in the shaded area of Fig.\ 6 or in the shaded area of Fig.\ 7 -- is our choice.  If we say the
particle is initially specified at time $t_0$, then it will occupy the shaded area of Fig.\ 6 when we understand
`simultaneous' to be the horizontal surface through $t_0$.  But it will occupy the shaded area of Fig.\ 7 when we understand
`simultaneous' to be the conic surface with $t_0$ at its vertex.

\section*{Appendix II}
It is possible to draw a Minkowski diagram using one vertex time $t$, with all the particles in the system mapped onto the surface
of its backward time cone.  We have seen that it is also possible to use a diagram with two or more vertices as will be done in the
present case of two independent particles.  

Consider a nucleus $n_1$ with a vertex $t_1$ as shown in the diagram of \mbox{Fig.\ 8a}, together with a second nucleus $n_2$
with a vertex $t_2$.  Both nuclei are radioactive, where the first decays at event \textbf{A} in Fig.\ 8b.  Multiple time
descriptions like this are possible only when the vertex events so identified have a space-like relationship to each other.  In this
case the vertices are chosen to \emph{follow} the separate particles, since there is no reason why the time scale must be the same as
that of the Minkowski observer.  The total wave function in Fig.\ 8a will be spread over both peaks.  It will be mapped onto the dashed
lines joined at their intersection as shown in the figure.  

\begin{figure}[b]
\centering
\includegraphics[scale=0.8]{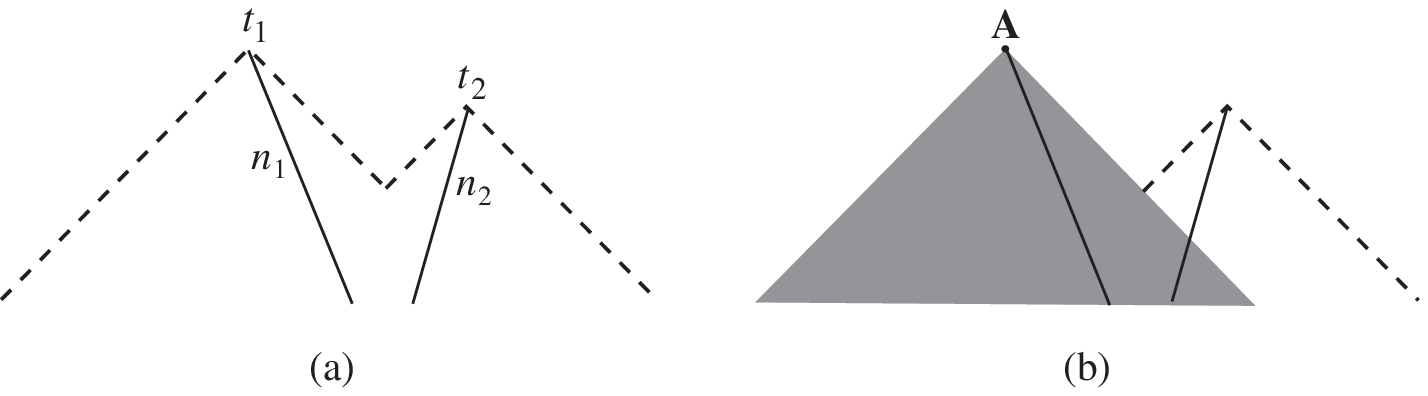}
\center{Figure 8: Two independent systems -- first reduction}
\end{figure}

The qRule equation follows Eq.\ 8 for these independent nuclei.  
\begin{displaymath}
U(t_1;t_2) = [n_1 + \underline{p}_1p_1'] [n_2 + \underline{p}_2p_2']
\end{displaymath}
where $n_1$ and $p_1p'_1$ are specified relative to the $t_1$ vertex, and $n_2$ and $p_2p'_2$ are specified relative to the $t_2$
vertex.  The component $p_1p'_1$ represents the decay products of $n_1$, and $p_2p'_2$ represents the decay products of $n_2$.  The
two ready components are initially zero.  Probability current will flow from both $n_1$ and $n_2$ in this equation to the
corresponding ready components, marking both candidates for a possible stochastic choice. 

The first ready component is stochastically chosen at Event \textbf{A}, after which the qRule equation is
\begin{equation}
U(t_1 = t_A;t_2) = p_1p_1'(n_2 + \underline{p}_2p_2')
\end{equation}
as shown in Fig.\ 8b.  At event \textbf{A} the first nucleus is reduced to its decay products, although they do not yet appear in
Fig.\ 8b.  The second nucleus in Eq.\ 11 continues unaffected in its unstable state, and is represented by the line that extends
beyond the shaded area.  That line is the same length in Fig.\ 8b as it is in \mbox{Fig.\ 8a} because it is not reduced by the first
decay.  

Further evolution of the system is described by same equation 
\begin{equation}
U(t_1 \ge t_A;t_2) = p_1p_1'(n_2 + \underline{p}_2p_2')
\end{equation}
except that the magnitude of the component $\underline{p}_2p'_2$ increases in time as represented by the longer line in Fig. 9a.  The
decay particles $p_1$ and $p_2$ are also shown in Fig.\ 9a.  They are actually waves spreading out in all directions from the vertex,
but they are shown as two distinct world lines as are all other particle paths in these diagrams.  The qRules do not specify paths,
only reduction events and their associated states.

With a stochastic hit on $\underline{p}_2p'_2$ in Eq.\ 12, the reduction at event \textbf{B} will yield
\begin{equation}
U(t_1 \ge t_A;t_2 =t_B) = p_1p_1'p_2p_2'
\end{equation}
This equation is shown in the Minkowski diagram of Fig.\ 9b, where the decay products $p_2 p'_2$ do not yet appear.   

\begin{figure}[b]
\centering
\includegraphics[scale=0.8]{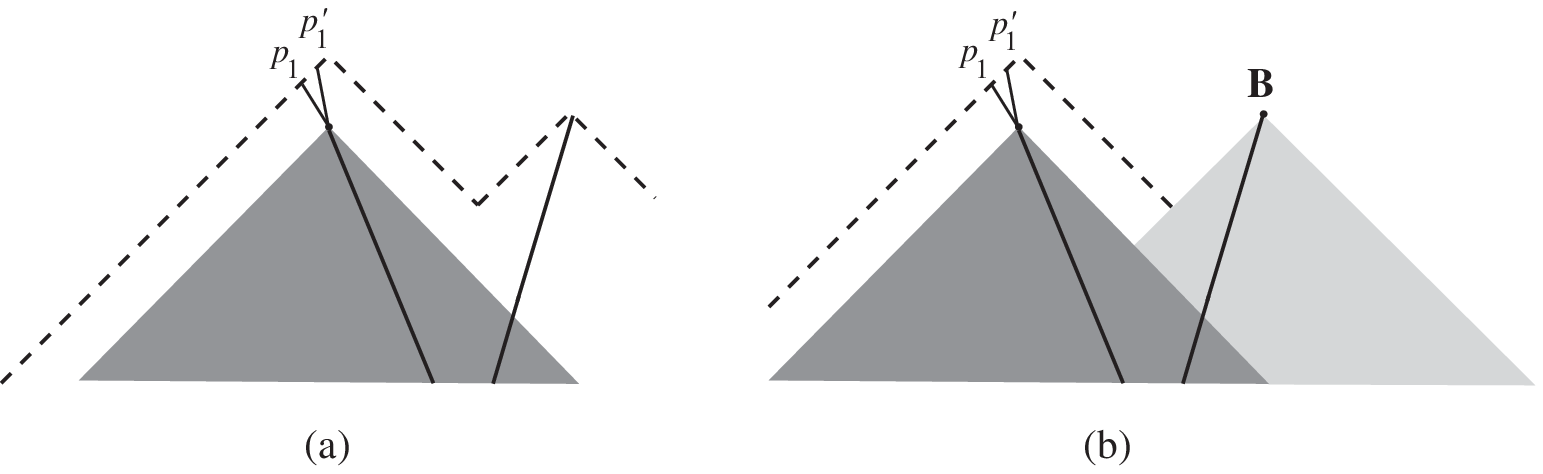}
\center{Figure 9: Two independent systems -- second reduction}
\end{figure}

As before, the lighter shaded reduction under event \textbf{B} in Fig.\ 9b does not penetrate the darker shaded reduction defined by
event \textbf{A}.  This is because the \textbf{B} reduction in  Eq.\ 13 occurs causally \emph{after} the \textbf{A}
reduction in Eq.\ 11, so it can have no influence on the evolution leading to Eq. 11.   

It was shown in the text that correlated reductions are causally sequenced so that a later reduction will not penetrate a former
reduction.  We see here that the same is true when the reductions are independent of one another.  A distinct causal order is
therefore characteristic of \emph{all} reductions.  As before, the reduction in \mbox{Fig.\ 8b} may appear to go back infinitely far
in time and extend infinitely far in space, but that will not happen.  Every reduction will find a floor of other mountaintops that
will support it and limit its range in space and time.

\end{document}